\documentclass[aps,twocolumn,prd,epsf]{revtex4}
\usepackage{amssymb}
\usepackage{graphicx}
\newcommand{\be}{\begin{equation}}
\newcommand{\ee}{\end{equation}}
\newcommand{\bea}{\begin{eqnarray}}
\newcommand{\eea}{\end{eqnarray}}

\newcommand{\lpl}{\ell_{\rm Pl}}

\begin{document}

\title{A graceful entrance to braneworld inflation}

\author{James E. Lidsey and David J. Mulryne}

\affiliation{Astronomy Unit, School of Mathematical Sciences,
Queen Mary, University of London, London E1 4NS, U.K.}

\begin{abstract}
Positively-curved, oscillatory universes have recently been shown
to have important consequences for the pre-inflationary dynamics of 
the early universe.  In particular, they may allow a self-interacting scalar
field to climb up its potential during a very large number of these
cycles. The cycles are naturally broken when the potential 
reaches a critical value and the universe begins to inflate, 
thereby providing a `graceful entrance' to early universe inflation. 
We study the dynamics of this behaviour within 
the context of braneworld scenarios which exhibit
a bounce from a collapsing phase to an expanding one. The dynamics can be
understood by studying a general class of braneworld models 
that are sourced by a scalar field  with a constant potential.
Within this context, we determine the conditions a 
given model must satisfy for a graceful entrance to be 
possible in principle. We consider the bouncing braneworld 
model proposed by Shtanov and Sahni and show that it exhibits 
the features needed to realise a graceful entrance to inflation
for a wide region of parameter space.

\end{abstract}
\maketitle

\section{Introduction}

The inflationary scenario is presently the favoured model 
for large-scale structure formation in the universe
\cite{inf1,inf2,inf3,inf4,inf5,linde83}. (For 
reviews, see \cite{llkcba,lythriotto,lidlyth,btw}). 
Inflation arises whenever the universe undergoes a phase of accelerated 
expansion. Within the context of relativistic 
cosmology, a necessary and sufficient condition 
for acceleration is a violation of the strong energy condition, 
i.e., $\rho +3 p< 0$. This is equivalent to the condition 
$\gamma < 2/3$, where $\gamma \equiv (\rho+ p)/\rho$
defines the equation of state parameter. 
In the simplest versions of the scenario, such a condition 
is realised by a scalar `inflaton' field that is rolling sufficiently slowly 
down its self-interaction potential, $\dot{\phi}^2 < V$ \cite{linde83}. 

Given the recent successes of inflation when confronted with 
observations of the Cosmic Microwave Background (CMB)
radiation and high redshift surveys \cite{wmap}, a key question to address 
is how the initial conditions for successful inflation were established
in the very early universe. This question was recently examined within
the context of Loop Quantum Cosmology (LQC), which is the application of 
Loop Quantum Gravity to symmetric states. (For reviews, see 
\cite{loopreview,loop1,bojoreview}). In this scenario, semi-classical 
modifications to the standard Friedmann equations 
cause a minimally coupled scalar field to 
behave as an effective phantom fluid below a critical 
value of the scale factor, $a_*$, with an effective equation of state, 
$\gamma_{\rm eff} < 0$ \cite{Bojowald2002,lmnt,mntl,mtle,nunes}. 
This causes a collapsing, isotropic, spatially closed 
Friedmann-Robertson-Walker (FRW) universe to undergo a non-singular bounce
into an expansionary phase. On the other hand, if the field satisfies the 
strong energy condition above $a_*$ (where classical dynamics is recovered), 
such a universe will eventually recollapse. The overall 
result, therefore, is that the universe oscillates between  
cycles of contraction and expansion \cite{lmnt}. 

If the field's kinetic energy dominates its potential energy
during these cycles, it will effectively behave as a massless 
field with its value either increasing or decreasing 
monotonically with time. In principle, therefore, the field may 
gradually roll {\em up} its potential (assuming implicitly that it is 
evolving along a region of the potential that is increasing in magnitude). 
As a result, the potential energy will gradually increase 
during each successive cycle, with the net result that it becomes progressively 
harder to satisfy the strong energy condition on scales 
above $a_*$. Eventually, 
therefore, a cycle will be reached when the strong energy condition 
becomes violated during the expansionary phase of the cycle
($a>a_*$), and this will initiate a phase of inflation. This implies that  
slow-roll inflation may be possible even if 
the field is initially located in a region of the potential 
that would not generate accelerated expansion, such as 
a minimum. This effect has
been termed a `graceful entrance' to inflation \cite{nunes}. 

A key feature of the dynamics described above is that 
the universe is able to oscillate because 
the effective equation of state satisfies $\gamma_{\rm eff} <
2/3$ for $a<a_*$ and $\gamma_{\rm eff} >2/3$ for $a>a_*$. 
Since these are rather weak conditions, it is important 
to investigate whether similar effects are possible in other 
cosmological scenarios, such as the braneworld paradigm. This 
is the purpose of the present paper. 
The braneworld scenario is motivated by string/M-theory 
\cite{string1,string2,string3} and
has attracted considerable attention in recent years. (See 
\cite{royreview,braxreview,braxreview1} for reviews). 
Our observable four-dimensional 
universe is interpreted as a co-dimension one brane 
propagating in a five- (or higher-) dimensional `bulk' space. 
A number of bouncing braneworld models have been developed to date
\cite{shsh,burgess,bounce1,bounce2,bounce3,bounce4}. 
In the model proposed by Shtanov and Sahni (S-S), 
for example, the extra bulk dimension is timelike and this results 
in modifications to the effective four-dimensional Friedmann equation 
that induce a non-singular bounce \cite{shsh}. 

The paper is organized as follows. In Section II, 
we outline how a phase space analysis for a field with a 
constant potential can yield valuable insight into the 
cosmic dynamics that leads to a graceful entrance for
inflation. We then proceed in Section III to determine the necessary and 
sufficient conditions for graceful entrance when the Friedmann 
equation has an arbitrary dependence on the 
energy density of the universe. We consider the S-S
braneworld scenario as an explicit example in Sections IV--VI and 
determine the regions of parameter space where a graceful entrance  
is possible. We conclude with a discussion in Section VII.

\section{The phase space description of graceful entrance
to inflation}

The oscillatory dynamics of isotropic universes 
sourced by a scalar field with a constant potential 
proves very useful when determining the necessary conditions 
for a graceful entrance to inflation for a given cosmological 
model. In general, the dynamics is determined by 
three evolution equations for the scale factor, the 
Hubble parameter and the velocity of the scalar field, respectively. 
In addition, there is the Friedmann equation,  which 
represents a constraint. This implies that the dynamics can be expressed 
as a two-dimensional system of equations and presented 
on a two-dimensional phase space. 

\begin{figure}[!t]
\includegraphics[width=6.05cm, height=6.05 cm]{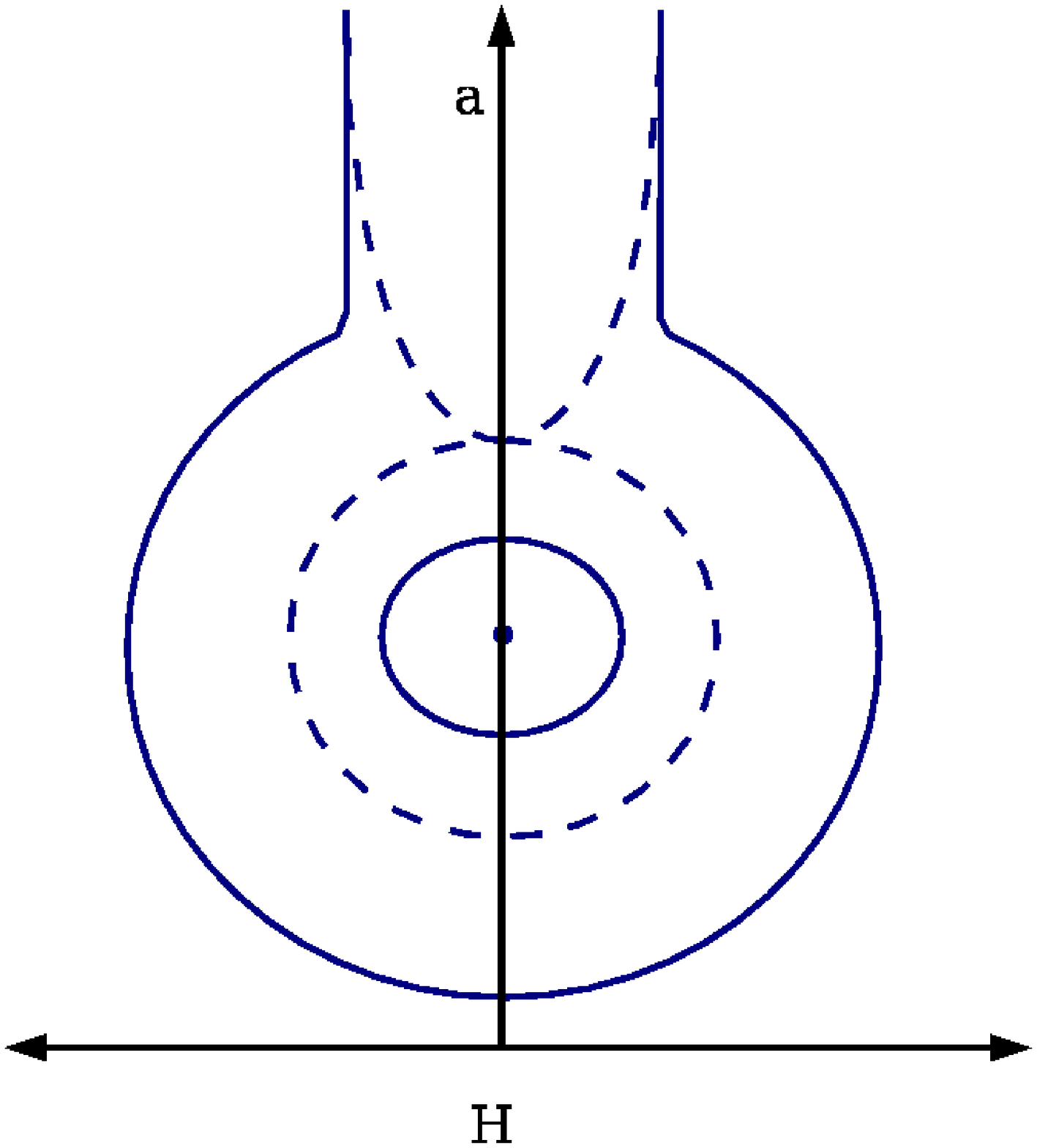}
\includegraphics[width=6.05cm, height=6.05 cm]{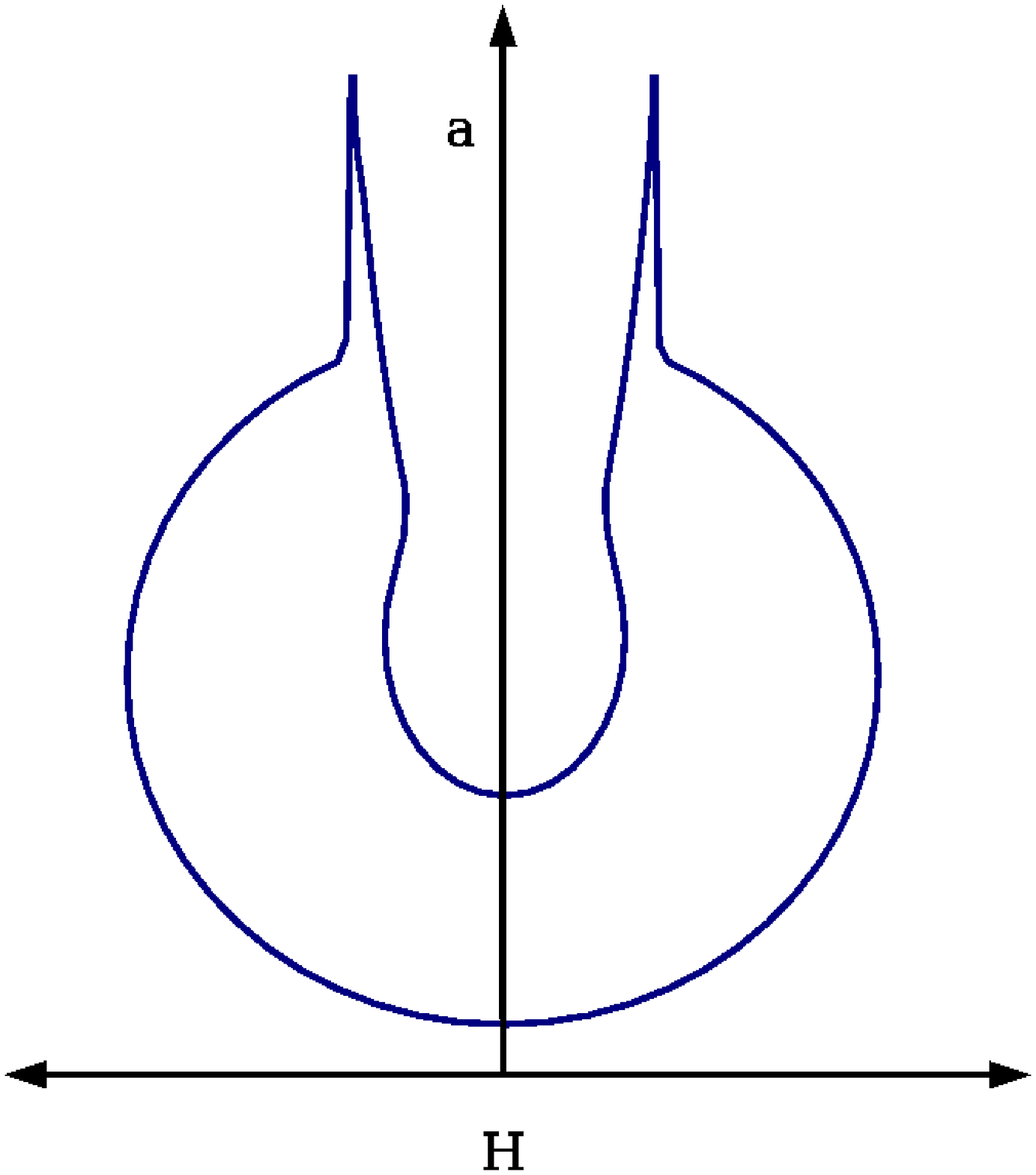}
\caption[] {\label{fig0}
Illustrating the general features of the phase space discussed in 
the text that are relevant to a graceful entrance to inflation.
The variables are $\{H,a\}$ and all trajectories evolve 
in an anti-clockwise direction. The top diagram corresponds to 
$0 <V< V_{\rm crit}$ and the bottom panel to $V>V_{\rm crit}$.
In the former case, the important features for a graceful entrance
are an area of phase space in which oscillations take place, as 
indicated by the region enclosed by the dotted line, 
and trajectories which evolve into an inflationary 
phase.  The centre equilibrium point about which oscillations occur 
is marked with a circle, the separatrix is represented by the dotted line 
and the saddle equilibrium point occurs at the self-intersection 
of the separatrix. In the lower panel $(V>V_{\rm crit})$, 
there exist no centre or saddle equilibrium points and 
all trajectories evolve into an inflationary phase.  
}
\end{figure}

In the LQC scenarios considered previously, 
it proved convenient to parametrize the phase space in terms of 
the scale factor and Hubble parameter \cite{mtle}. 
The qualitative nature of the phase space that can result in a
graceful entrance to inflation is illustrated in Fig. \ref{fig0}. 
For a positive potential taking values in the range
$0 <V< V_{\rm crit}$, where $V_{\rm crit}$ is some critical 
value that is determined by the parameters of 
the model, the phase space contains a 
centre and a saddle equilibrium point. Two types of 
behavior are therefore possible. Trajectories that 
are sufficiently close to the centre encircle it 
and represent trajectories that undergo eternal 
oscillations. On the other hand, trajectories which pass above the saddle 
point represent initially collapsing universes which evolve 
through a bounce into an eternal (de Sitter) inflationary era. 
This is illustrated qualitatively in the top panel of Fig. \ref{fig0}. 

If the value of the potential is increased, 
the effect on the phase space is such that the centre and saddle 
point are positioned closer to one another. At the critical 
value $V=V_{\rm crit}$, the points merge and disappear. 
The phase space for $V>V_{\rm crit}$ therefore contains 
no equilibrium points and all trajectories represent initially contracting  
universes which bounce and subsequently inflate, as shown in the lower
panel of Fig. \ref{fig0}.

This simplified phase space structure enables us to 
develop a graceful entrance mechanism for more 
realistic cosmologies, where the potential is field-dependent. 
Let us assume that the scalar field is initially located in a region of the 
potential such that $V(\phi ) <V_{\rm crit}$ and that the universe 
is undergoing oscillations about the centre equilibrium point. 
The dynamics in this regime will be dominated by the field's kinetic energy and 
we will further suppose that the field's velocity is such that it moves 
up the potential. If the change in the 
magnitude of the potential is negligible over 
a given cycle, this cycle can still be represented as a trajectory 
in the phase space associated with an instantaneous value of the 
potential, as illustrated in the top panel of Fig. \ref{fig0}.  
Over a large number of cycles, however, this `instantaneous 
phase space' will become significantly modified. Indeed, if 
the potential energy increases monotonically as the field evolves,  
the region of phase space in which oscillations are possible will
become progressively smaller.  Since the oscillatory region 
disappears completely at $V=V_{\rm crit}$, this corresponds to the 
largest value that the field's potential energy can acquire before 
the oscillations are broken. At this point the trajectory 
in the phase space rapidly evolves into the inflationary 
regime corresponding to the lower panel of Fig. \ref{fig0}. 
Hence, the value $V_{\rm crit}$ 
sets the energy scale at the onset of inflation. 
Once inflation begins, the field's kinetic energy will 
rapidly tend to zero as the field slows down, reaches a point of maximum 
displacement and rolls back down the potential. Slow-roll 
inflation will then arise if the 
potential satisfies the usual slow-roll constraints.

Although the dynamics may be further complicated by the introduction 
of additional matter sources \cite{nunes}, the above description outlines how 
a field can evolve up its self-interaction potential 
while the universe undergoes oscillations. 
This forms the basis for a graceful entrance mechanism that 
naturally generates 
the conditions for slow-roll inflation. A phase plane analysis is 
important since it highlights the relevance of the centre and 
saddle equilibrium points. In particular, a sufficient but not necessary 
condition for cyclic behaviour is the existence of 
a centre equilibrium point, while a saddle point 
is required to separate those regions of phase space where 
cyclic and non-cyclic behaviour takes place. Furthermore, 
the disappearance of the saddle point at a critical value of 
the potential is key to the graceful entrance mechanism. 

In the following Section, therefore, we will consider 
the conditions for the existence of saddle and centre equilibrium 
points in cosmologies described by a set 
of generalized Friedmann equations. 

\section{General conditions for centre and saddle equilibrium points}

\subsection{Relativistic Cosmology} 

We wish to study classes of cosmological models that are motivated 
by the braneworld paradigm and to determine 
whether such models display the general characteristics required for  
a graceful entrance to inflation. However, we will first consider 
the relativistic cosmology based on classical Einstein gravity, since 
this system provides a suitable framework for considering 
more general models. Specifically, we will consider a 
positively-curved FRW cosmology 
sourced by an effective perfect fluid with an energy density 
and pressure related by an arbitrary equation 
of state, $p_{\rm eff} = [\gamma_{\rm eff} -1] \rho_{\rm eff}$, where
it is assumed implicitly that  
the equation of state parameter is a known 
function of the scale factor, $\gamma_{\rm eff} = \gamma_{\rm eff}
(a)$. 

The Friedmann and fluid equations for this model are given by 
\bea 
\label{GenFried}
H^2 = \frac{8 \pi \lpl^2}{3}\rho_{\rm eff}-\frac{1}{a^2}\,~, \\
\label{GenEnCon}
\dot{\rho}_{\rm eff}=-3H\gamma_{\rm eff}\rho_{\rm eff}\,~,  
\eea
where
\be
\label{2DSysGen}
\dot{a}=Ha\,~,
\ee
and Eqs. (\ref{GenFried}) and (\ref{GenEnCon}) 
fully determine the cosmic dynamics. Differentiating 
Eq. (\ref{GenFried}) with respect to cosmic time implies that 
\be
\label{HdotGen}
\dot{H}=-\frac{3 \gamma_{\rm eff}}{2} H^2 
+  \left (1-\frac{3 \gamma_{\rm eff}}{2}\right )
\frac{1}{a^2}\,~,
\ee
and Eqs. (\ref{2DSysGen}) and 
(\ref{HdotGen}) then describe a closed dynamical system, where the 
Friedmann equation (\ref{GenFried}) represents a constraint. 
The equilibrium points of this system arise whenever 
$\dot{a}=\ddot{a}=0$, which implies that
\be
\label{FixCond}
\gamma_{\rm eff} (a_{\rm eq}) =\frac{2}{3}\,~, \qquad   H (a_{\rm eq}) 
=0\,~.
\ee

The stability of the system (\ref{2DSysGen})-(\ref{HdotGen}) 
can be determined by linearising about the equilibrium points 
and evaluating the corresponding eigenvalues. It is straightforward 
to show that the eigenvalues are given by 
\be
\label{EigenValsGen}
\lambda^2 = -\frac{3}{2} \left[ \frac{1}{a^2} 
\frac{d \gamma _{\rm eff}}{d \ln  a} \right]_{\rm eq}\,~.
\ee
In general, therefore, a necessary and sufficient 
condition for the equilibrium point to be a centre is
\be
\label{genpointcentre}
\lambda^2 <0\,~, \qquad  H=0\,~,\qquad 
\gamma_{\rm eff}=\frac{2}{3}\,~, \qquad 
\frac{d \gamma_{\rm eff}}{d \ln a}>0\,~,
\ee
whereas the corresponding condition to be a saddle is 
\be
\label{genpointsaddle}
\lambda^2  >0\,~,\qquad H=0\,~, \qquad
\gamma_{\rm eff}= \frac{2}{3}\,~, \qquad 
\frac{d \gamma_{\rm eff}}{d \ln a}<0\,~.
\ee
It is interesting to note that under the assumptions we have 
made, centre and saddle points are the only types of equilibrium 
points permitted.

\subsection{Braneworld Scenarios} 

The four-dimensional cosmological dynamics for a wide class of 
positively-curved FRW braneworld 
scenarios can be modeled in terms of a generalized 
Friedmann equation of the form 
\be 
\label{GenBraneFried}
H^2 = \frac{8 \pi \lpl^2}{3}\rho L^2 (\rho) 
+f(a)-\frac{1}{a^2}\,~,
\ee
where $\rho$ is the total energy density of the matter confined 
to the brane. The function $L (\rho )$ is assumed to be positive-definite
and parametrizes the departure of the model from the 
standard relativistic  behaviour, $H^2 \propto \rho$. The 
function $f(a)$ is a function of the scale factor and, in a braneworld 
context, usually parametrizes the effects of a bulk black hole 
on the four-dimensional dynamics \cite{SMS,muk,ida}. 
If the matter fields are confined 
to the brane, they satisfy the standard conservation equation 
\be
\label{BraneEnCon}
\dot{\rho}=-3H\gamma \rho\,~,
\ee
where the equation of state parameter is defined by 
$p=(\gamma-1)\rho$ and is once more to be viewed 
implicitly as a known function of the scale factor. 

The system (\ref{GenBraneFried})-(\ref{BraneEnCon}) can be 
expressed in the standard form of Eqs. (\ref{GenFried})-(\ref{GenEnCon}) 
by defining an effective energy density \cite{clm}
\be
\label{rhoEffBrane}
\rho_{\rm eff}=\rho L^2 + \frac{3}{8 \pi \lpl^2} f(a)\,~.
\ee
Differentiating Eq. (\ref{rhoEffBrane}) and 
comparing with Eq. (\ref{GenEnCon}) then
implies that the effective equation of state 
parameter has the form 
\begin{eqnarray}
\label{gammaGen}
\gamma_{\rm eff}= \left[ \gamma \left( \rho L^2 +\rho^2 \frac{d(L^2)}{d
\rho} \right) - \frac{1}{8 \pi \lpl^2} \frac{df}{d \ln a} \right]
\nonumber \\
\times \left( \rho L^2 + \frac{3}{8\pi \lpl^2} f \right)^{-1}\,~.
\end{eqnarray}

In principle, therefore, if the conservation 
equation (\ref{BraneEnCon}) can be solved for 
a given equation of state, $\gamma (a)$, the effective equation 
of state (\ref{gammaGen}) will be a known function of the 
scale factor once the functional forms of $L(\rho )$ and $f(a)$ 
have been specified for a particular braneworld model. 
Differentiation will then yield the necessary information 
required in order to determine 
the existence (or not) of the centre and saddle equilibrium points, 
as summarized in Eqs. (\ref{genpointcentre}) and (\ref{genpointsaddle}). 

In the following Section, we will employ these results 
within the context of a specific braneworld model. 

\section{Shtanov-Sahni Braneworld}

The Shtanov-Sahni (S-S) braneworld scenario embeds 
a co-dimension one brane with a negative tension $\tilde{\sigma}$ 
in a five-dimensional Einstein space sourced by a 
positive cosmological constant, where the fifth 
dimension is timelike \cite{shtanov,shsh}. 
The effective Friedmann equation on the brane is given by 
\cite{shtanov,shsh,kofinas}
\be
\label{FriedSahni}
H^2=\frac{8\pi \lpl^2}{3} \left [\rho - \frac{\rho^2}{2\sigma} 
\right] + \frac{m}{a^4} - \frac{1}{a^2}\,~,
\ee
where $\sigma \equiv -\tilde{\sigma}$. If the bulk space 
is not conformally flat, the projection of the five-dimensional 
Weyl tensor induces an effective `dark radiation' term on 
the brane, parametrized by the constant $m$ \cite{SMS,muk,ida}. 
A conformally flat bulk corresponds to $m=0$. 
Under quite general conditions, a negative quadratic dependence on 
the energy density in the Friedmann equation (\ref{FriedSahni}) implies 
that a collapsing brane world-volume is able to undergo a non-singular bounce
\cite{shsh}.

Comparison of the Friedmann equations (\ref{GenBraneFried}) 
and (\ref{FriedSahni}) implies immediately that 
\be
\label{compare}
L^2= \left (1- \frac{\rho}{2 \sigma} \right)\,~,~~
f(a)=\frac{m}{a^4}\,~,
\ee
and substituting Eq. (\ref{compare}) into Eq. (\ref{gammaGen}) implies that 
the effective equation of state parameter is given by 
\be
\label{gammasahni}
\gamma_{\rm eff} = \frac{ \gamma \rho( 1 - \rho / \sigma) a^4 + 
m/2\pi \lpl^2}{\rho(1-\rho/2\sigma) a^4  + 3m/8 \pi \lpl^2}\,~.
\ee

Since we are interested in whether a graceful 
entrance to inflation can occur in this model, 
we will determine the equilibrium points that arise when 
the matter confined to the brane corresponds to a 
scalar field that is rolling along a constant 
potential, $V$. This is equivalent to a matter 
source comprised of a massless scalar field and a cosmological 
constant. The equation of state for the system is given by 
\begin{equation}
\label{eosfield}
\gamma = 2 \left( 1-\frac{V}{\rho} \right)\,~.
\end{equation}

In general, the equilibrium points for 
this system will arise whenever Eqs. (\ref{genpointcentre}) or 
(\ref{genpointsaddle}) are satisfied. 
It follows from Eqs. (\ref{FriedSahni}) 
and (\ref{gammasahni}) that such points 
occur when 
\bea
\label{FriedSahniZero}
\frac{8 \pi \lpl^2}{3} \left ( \rho-\frac{\rho^2}{2\sigma} \right ) 
+ \frac{m}{a^4} = \frac{1}{a^2}\,~, \\
\label{gammasahniFix}
\gamma\left (\rho - \frac{\rho^2}{\sigma}\right ) 
- \frac{2}{3}\left ( \rho -\frac{\rho^2}{2\sigma} \right)
= -\frac{m}{4 \pi \lpl^2} \frac{1}{a^4}\,~.
\eea
For finite values of the scale factor, 
Eq. (\ref{FriedSahniZero}) may be simplified after 
substitution of Eq. (\ref{gammasahniFix}):
\be
\label{asquared}
\frac{1}{a^2}=4 \pi \lpl^2  \rho \left [ \frac{4}{3} -
\frac{2\rho}{3\sigma} + \gamma \left (\frac{\rho}{\sigma}-1 \right ) 
\right ]\,~,
\ee
and Eq. (\ref{asquared}) may then be employed to 
express Eq. (\ref{FriedSahniZero}) in the form of 
a quartic equation in the energy density: 
\begin{eqnarray}
\label{quartic}
 8 \pi \lpl^2 m \left[ 
\frac{2\rho^2_{\rm eq}}{3\sigma}-
\left(\frac{1}{3}+\frac{V}{\sigma}\right)\rho_{\rm eq} +V \right]^2
\nonumber \\
- \frac{5\rho^2_{\rm eq}}{6\sigma} + \left ( \frac{2}{3}+\frac{V}{\sigma} 
\right)\rho_{\rm eq}  -V =0\,~.
\end{eqnarray}
The solutions to Eq. (\ref{quartic}) yield the values of the energy 
density at the equilibrium points and the corresponding value 
of the scale factor can then be deduced directly 
from Eq. (\ref{asquared}). For physical solutions, one must 
ensure that the energy density 
and scale factor are positive and real at each equilibrium point.  

The dark radiation on the brane can significantly 
influence the dynamics of the system. In view of this, 
we consider separately the cases where this radiation is 
present or absent in the following Sections. 

\section{No Dark Radiation}

If no dark radiation is present ($m=0$), 
the quartic equation (\ref{quartic}) reduces to the quadratic constraint
\be
\label{quadratic}
\frac{10\rho^2_{\rm eq}}{3\sigma}-\left ( \frac{8}{3}+\frac{4V}{\sigma}
\right)\rho_{\rm eq} +4V = 0\,~, 
\ee
and this equation can be solved in terms of the field's kinetic energy:   
\be
\label{roots}
\dot{\phi}^2_{\rm eq}= \frac{4 \sigma - 4 V \pm 2 \left [
9V^2 - 18 V \sigma +4 \sigma^2\right ]^{1/2}}{5}\,~.
\ee
It follows that there can be at most 
two static (physical) solutions to the Friedmann equations
(\ref{GenBraneFried}) and (\ref{BraneEnCon}), depending on the 
magnitude of the potential. If $V <0$, 
there is only one solution to Eq. (\ref{quadratic})
with $\dot{\phi}^2>0$, implying there is only one equilibrium point 
in the phase space. For $V=0$, there is 
one solution with $\dot{\phi}^2 >0$ and a second, but physically 
uninteresting, point where the field's kinetic energy vanishes and 
the scale factor diverges. For a positive 
potential, on the other hand, there are two real roots to Eq. 
(\ref{quadratic}) if the condition
\be
\label{mergequad}
9V^2-18V\sigma+4\sigma^2>0\,~,
\ee
is satisfied. 
The values of the potential which bound the region of parameter 
space in which there are 
no real solutions are therefore given by 
\be
\label{Vcrit}
V =\left ( 1 \pm \frac{\sqrt{5}}{3} \right ) \sigma\,~.
\ee
We find that for $0< V < V_{\rm crit}$, where  
\begin{equation}
\label{Vcritdef}
V_{\rm crit} \equiv 
\left ( 1 - \frac{\sqrt{5}}{3} \right ) \sigma\,~,
\end{equation}
there are two real solutions 
to Eq. (\ref{quadratic}). Moreover the roots are 
physical, since they are positive 
and lead to real values of the 
scale factor. The roots merge and disappear at $V_{\rm crit}$.
Although other solutions to Eq. (\ref{quadratic}) arise 
when the potential is greater than the positive branch 
of Eq. (\ref{Vcrit}), these do not 
correspond to physical equilibrium points, either because 
the solution requires $\dot{\phi^2} <0$ or the 
value of the scale factor is imaginary.

Hence, there are two static equilibrium 
points for $0< V < V_{\rm crit}$. The nature of these points can be determined 
from condition (\ref{EigenValsGen}) after substituting 
Eqs. (\ref{gammasahni}), (\ref{FriedSahniZero}) and (\ref{quadratic})
and we find that the eigenvalues are given by 
\be
\label{eigenvalues}
\lambda^2 = 4 \pi \lpl^2 \dot{\phi}^2 \left[ 4- \frac{1}{\sigma} 
\left( 5 \dot{\phi}^2 + 4V \right) \right]\,~.
\end{equation}
This implies that a given point will correspond 
to a centre if 
$\dot{\phi}^2 > \frac{4}{5}\left(\sigma-V\right)$, 
otherwise it will represent a saddle.  Consequently, for
physical roots to Eq. (\ref{quadratic}), the field's kinetic 
energy has a value
\be
\label{centre}
\dot{\phi}_{\rm eq}^2=\frac{-4V+4\sigma 
+ 2 \sqrt{9V^2 - 18V\sigma + 4\sigma^2}}{5}\,~,
\ee
at the centre equilibrium point, whereas it takes the value 
\be
\label{saddle}
\dot{\phi}_{\rm eq}^2=\frac{-4V+4\sigma - 
2 \sqrt{9V^2 - 18V\sigma + \sigma^2}}{5}\,~,
\ee
at the saddle. 

It follows from Eqs. (\ref{centre}) and (\ref{saddle}) that 
the centre and saddle equilibrium points move 
towards each other as the value of the potential is increased, 
eventually merging and disappearing at $V=V_{\rm crit}$. 
These are precisely the requirements for a graceful entrance to inflation 
that were outlined in Section II.  

We now proceed to illustrate the cosmic dynamics in the phase 
space. The dynamics is clearly two-dimensional due to the Friedmann 
constraint equation (\ref{FriedSahni}). In principle, therefore, 
the phase space could be represented in two dimensions by 
introducing appropriate variables. However, this may 
be non-trivial in practice, since more than one set of variables 
may be required. (This would then imply that 
more than one phase plane diagram would be needed in order 
to present the full dynamics). 
In the context of the present discussion, it proves   
convenient to numerically integrate the field equations and then 
view the evolution of the universe in the three-dimensional 
space spanned by $\{ a, H, \dot{\phi} \}$. Since the Friedmann constraint  
defines a surface in this phase space,  
the evolution of the universe can then be parametrized in terms 
of a trajectory on this surface. In the following, 
we will refer to such a surface as the Friedmann surface. 
Identifying a coordinate system that covers  
the Friedmann surface is then equivalent to 
finding variables in which the dynamical 
system becomes explicitly two-dimensional, and  
the need for more than one coordinate patch to cover the surface 
is then equivalent to requiring more than one set of variables 
to complete the phase space.  

\begin{figure}[!t]
\includegraphics[width=8.05cm, height=6.6cm]{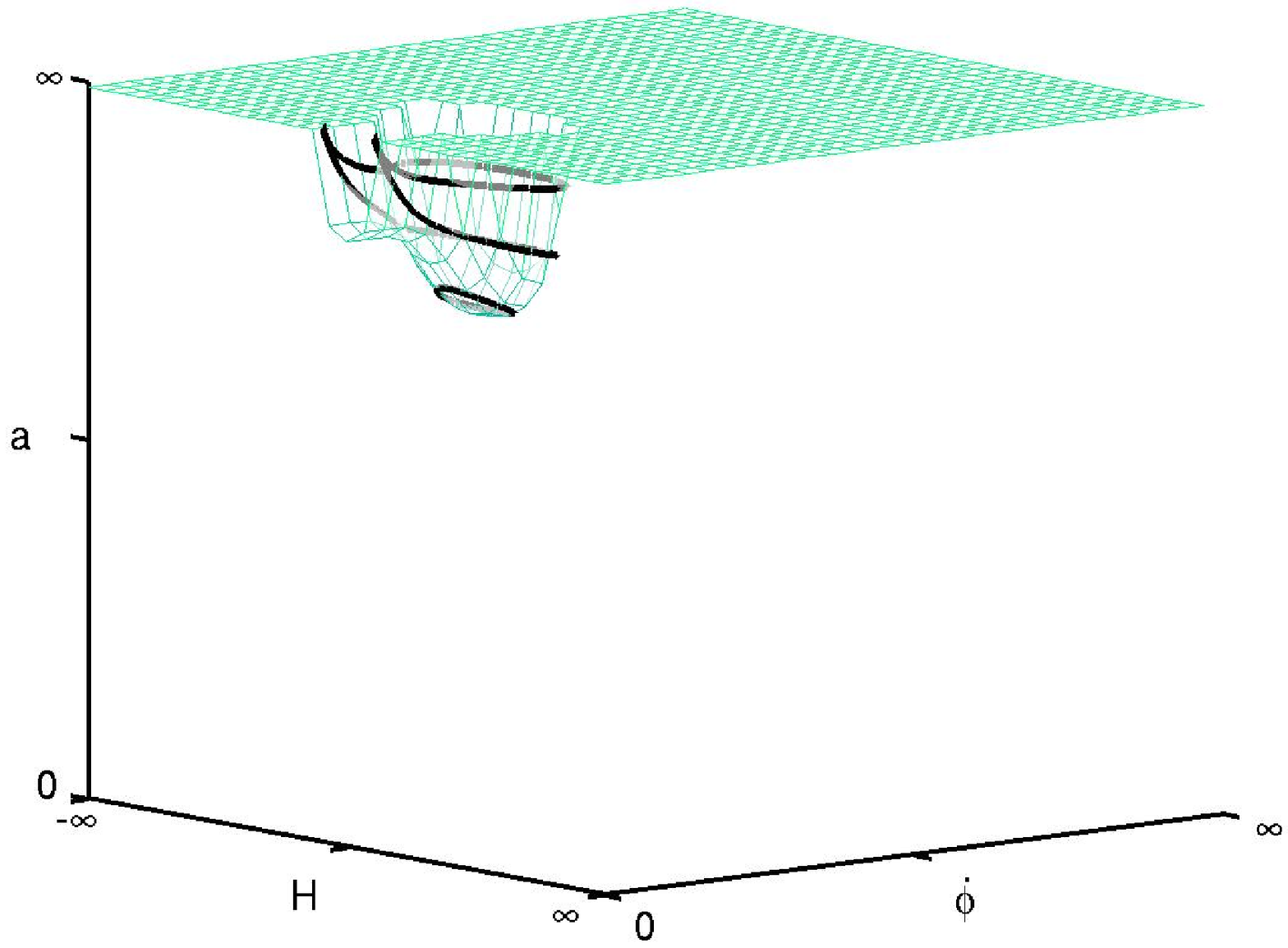}
\includegraphics[width=7.2cm]{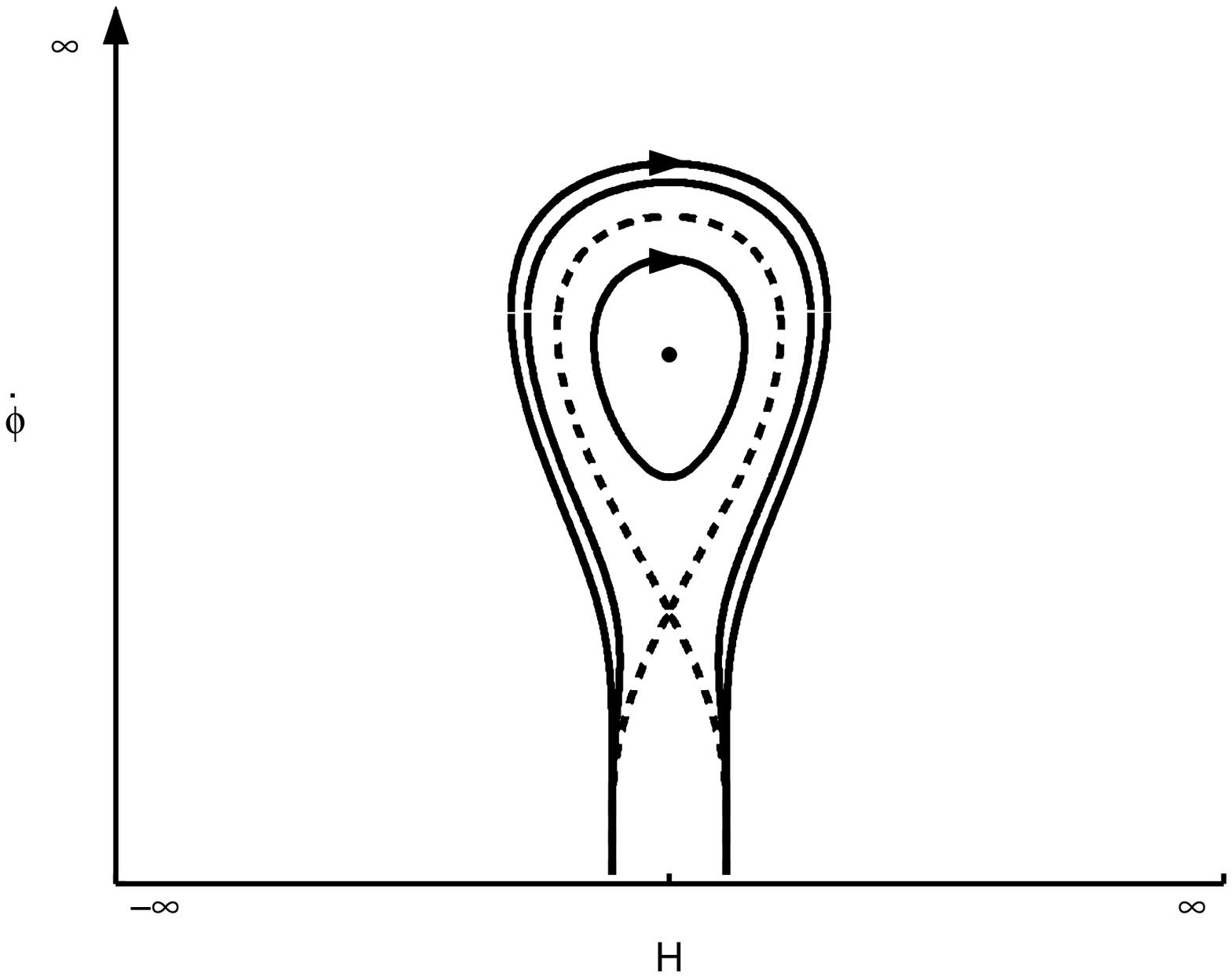}
\caption[] {\label{fig1}
The top panel represents the phase space in the variables
$\{H,\dot{\phi},a\}$, when
$m=0$, $\sigma=0.05$ and $0<V<V_{\rm crit}$, where $V_{\rm crit}$ is defined in 
Eq. (\ref{Vcritdef}). Numerical values are given in Planck units. The
surface defined by the Friedmann equation (\ref{FriedSahni}), with
$\rho=\dot{\phi}^2/2+V$, is also shown and all phase space
trajectories lie on this surface.  The compactified coordinates
$x \equiv {\rm arctan} (H)$, $y \equiv {\rm arctan}(\ln \dot{\phi})$ and 
$z \equiv {\rm arctan}(\ln a)$ have been employed 
in order to show the entire phase space.
The lower panel represents the two-dimensional projection of this
space onto the plane spanned by the variables $\{H,\dot{\phi}\}$.
The centre equilibrium point is denoted by the solid circle and the
separatrix is represented by the dotted line. The saddle point occurs
at the point where the separatrix self-intersects. The
axes have been compactified using the coordinate 
change $x={\rm arctan} (H)$ and $y= {\rm
arctan}(\ln \dot{\phi})$.
}
\end{figure}

\begin{figure}[!t]
\includegraphics[width=8.1cm, height=6.6cm]{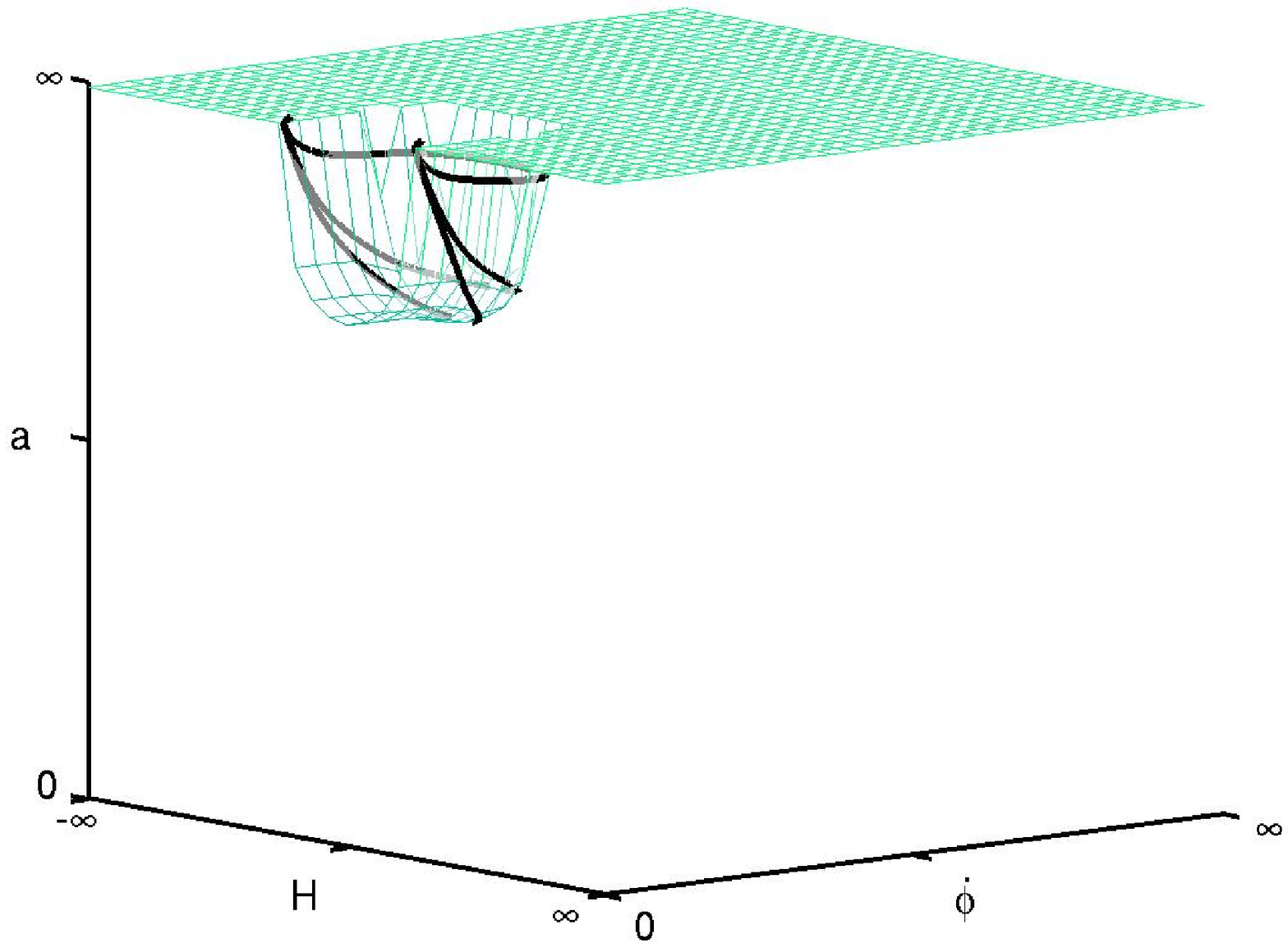}
\includegraphics[width=7.2cm]{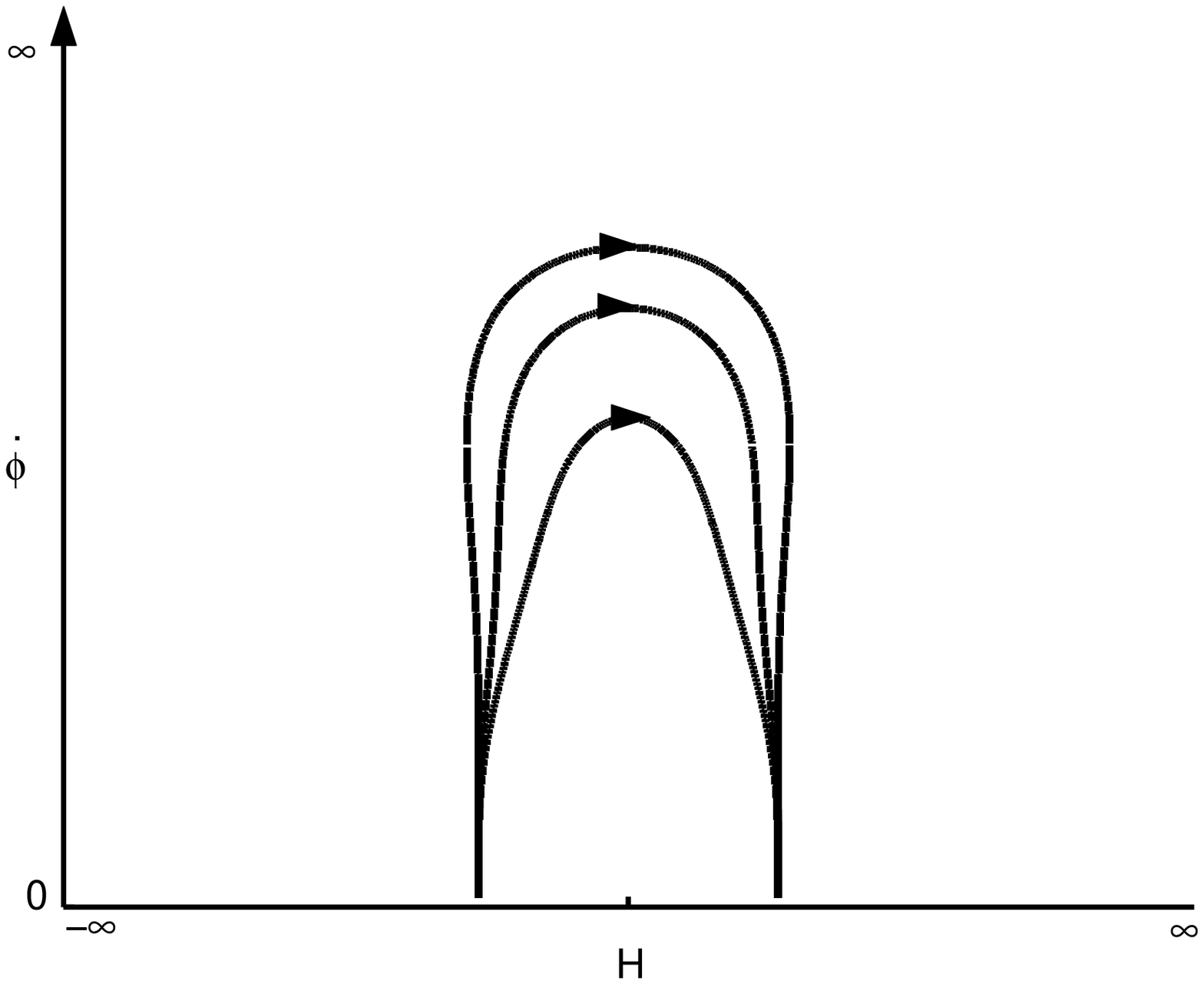}
\caption[] {\label{fig2}
As for Fig. \ref{fig1}, but now for a potential $V>V_{\rm crit}$.  
There are no finite equilibrium points in the phase space in this 
region of parameter space. 
}
\end{figure}

We focus on the case of a positive potential since we are interested 
in establishing the conditions for inflation via the graceful 
entrance mechanism. Fig. \ref{fig1} illustrates the Friedmann surface and the 
integrated trajectories on this surface when $V<V_{\rm crit}$. 
The lower part of the figure represents 
the projection of these trajectories onto the 
two-dimensional plane spanned by the variables  
$\{\dot{\phi},H\}$. The two equilibrium points representing the centre 
and saddle are shown. Fig. \ref{fig2}
illustrates the corresponding dynamics for $V>V_{\rm crit}$, 
where the equilibrium points have effectively merged. 
{\em All} trajectories now evolve through a bounce into an 
ever-expanding inflationary period of de Sitter expansion 
with $H \rightarrow {\rm constant}$, $\dot{\phi} \rightarrow 0$
and $a \rightarrow \infty$. 

Comparison between the phase spaces of 
Fig. \ref{fig0}  and those of Figs. \ref{fig1} and \ref{fig2} 
confirms that a graceful entrance to inflation, as 
outlined in section II, can occur in this model. 


\section{Effects of Dark Radiation}

In the case where dark radiation plays a dynamical 
role in the Friedmann equation (\ref{FriedSahni}), 
the full quartic expression given by Eq. (\ref{quartic}) 
must be solved to determine the nature of the equilibrium points.  
One must also ensure that the scale factor and kinetic energy 
of the field take positive and real values at these points. 
The constraint (\ref{quartic}) can be solved 
analytically using the standard techniques for 
quartic equations, but since the resulting expressions for the 
roots are very lengthy, we will not present them here. 
However, further insight may be gained by investigating 
the topology of the Friedmann surface. This may be achieved  
by rewriting the Friedmann constraint equation (\ref{FriedSahni}) 
in the form of a hyperboloid: 
\begin{eqnarray}
\label{FriedHy}
H^2 + \frac{8\pi\lpl^2}{3} \left (\frac{\dot{\phi}^2}{2 \sqrt{2 \sigma}} - \left (
\frac{\sqrt{2 \sigma}}{2} -\frac{V}{\sqrt{2 \sigma}} \right ) \right )^2 
\nonumber \\ 
- \left ( 
\frac{\sqrt{m}}{a^2}-\frac{1}{2\sqrt{m}}\right)^2 
= \frac{4 \pi \lpl^2 \sigma}{3} - \frac{1}{4 m}\,~.
\end{eqnarray}
Provided $V<\sigma$, 
the hyperboloid's topology depends only 
the relative values of the dark radiation parameter, $m$, 
and the brane tension, $\sigma$.
Indeed, there is a critical value of $m$ at which the topology 
changes and this is given by 
\begin{equation}
\label{mcrit}
m_{\rm crit} = \frac{3}{16 \pi \lpl^2} \frac{1}{\sigma}\,~.
\end{equation}
Consequently, the surface defined by Eq. (\ref{FriedHy}) 
consists of two 
disconnected pieces if $m < m_{\rm crit}$, whereas it is a single 
surface for $m> m_{\rm crit}$. 
It is possible that one of the two surfaces 
may correspond to a region where $\dot{\phi}^2<0$, 
in which case it would be unphysical. However, the 
qualitative dynamics of the universe will clearly be radically 
different above and below the critical value $m_{\rm crit}$
and we therefore proceed to consider each regime in turn.  

\subsection{$m<m_{\rm crit}$}

Since the analytic expressions for the solutions to the quartic 
equation (\ref{quartic}) are not particularly illuminating, 
we have employed them graphically to illustrate 
how the nature of the equilibrium points 
depends on the field's kinetic and potential energies for 
specific values of the dark radiation parameter and the brane tension.
Eq. (\ref{asquared}) was also employed to verify that the solutions to 
Eq. (\ref{quartic}) correspond to physical values of the scale factor,
and their stability was determined from  Eq. (\ref{EigenValsGen}). 

\begin{figure}[!t]
\includegraphics[width=8.2cm]{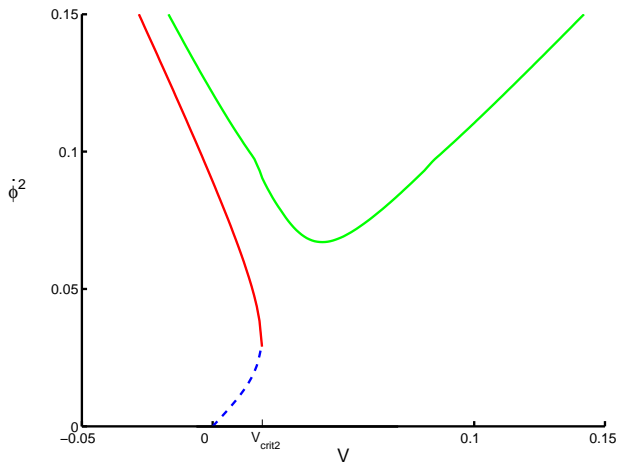}
\caption[] {\label{fig3}
Illustrating the position and nature of 
the equilibrium points as a function of the field's kinetic 
and potential energies when $0<m<m_{\rm crit}$. 
A dashed line corresponds to a saddle point and a solid line to a centre 
point. The critical value $V_{\rm crit2}$ represents the magnitude of 
the potential when one of the centre points and the saddle disappear 
from the phase space. The numerical values chosen for the plot
were $\sigma=0.05$ and $m=1.0$ and the axes are measured in Planck units.  
}
\end{figure}

The results are shown in Fig. \ref{fig3} for $m=1$ and $\sigma =0.05$, 
where the locations of the saddle and centre equilibrium points 
for given field energies are represented by the
solid and dashed lines, respectively. 
The qualitative behaviour is the same for all $m < m_{\rm crit}$.
There is a critical value of the potential, $V_{\rm crit2}$, 
such that there are three real roots to 
the quartic equation (\ref{quartic}) when $0< V<V_{\rm crit2}$.
These correspond to three distinct equilibrium points, 
two of which are centres whereas the third is a saddle point. 
At $V=V_{\rm crit2}$, however, the saddle and one of the centre points 
merge and, for $V>V_{\rm crit2}$, these points correspond to 
unphysical roots. Consequently, there is only one 
equilibrium point for $V>V_{\rm crit2}$ and this is a centre. 

\begin{figure}[!t]
\includegraphics[width=8.05cm, height=6.6cm]{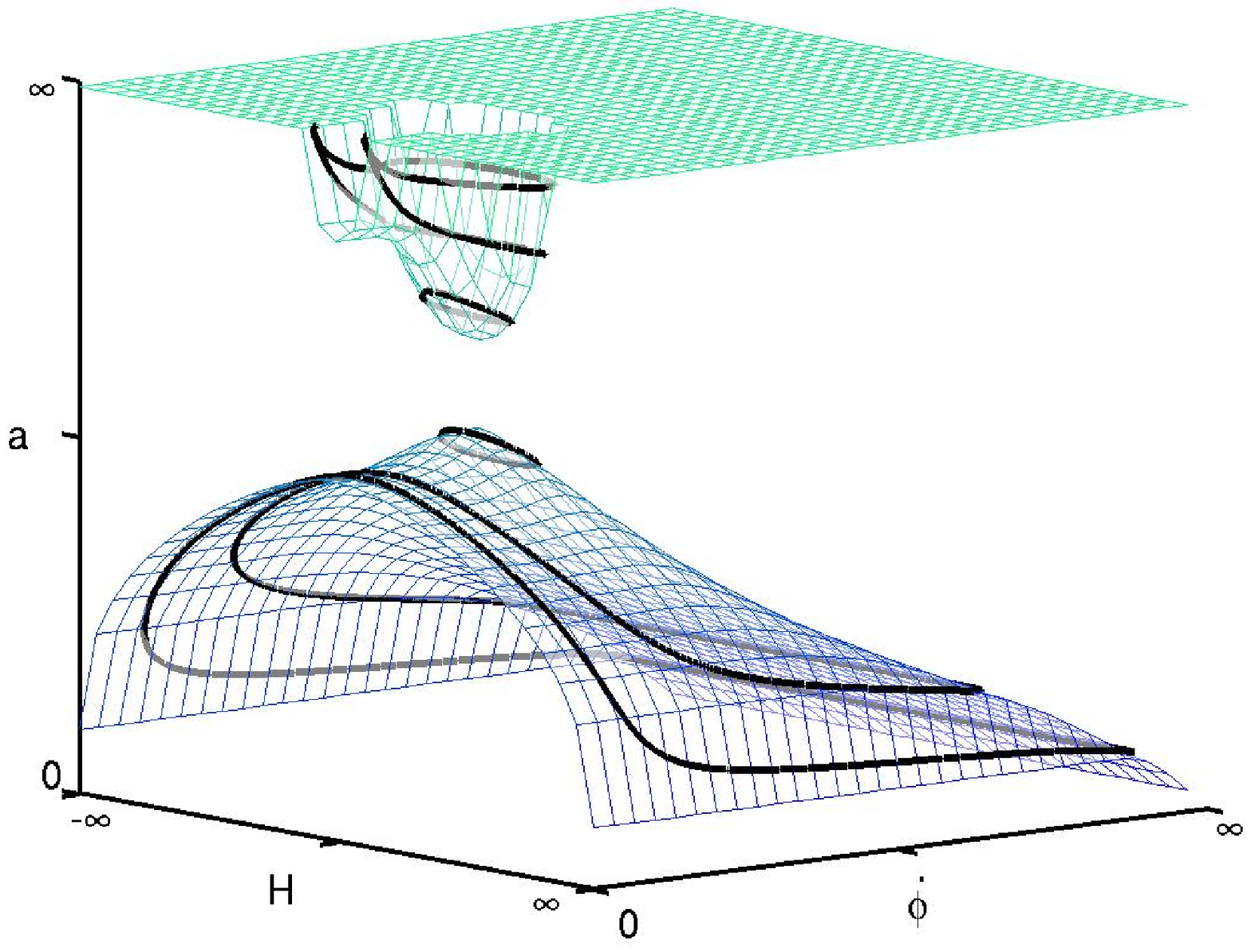}
\includegraphics[width=7.2cm]{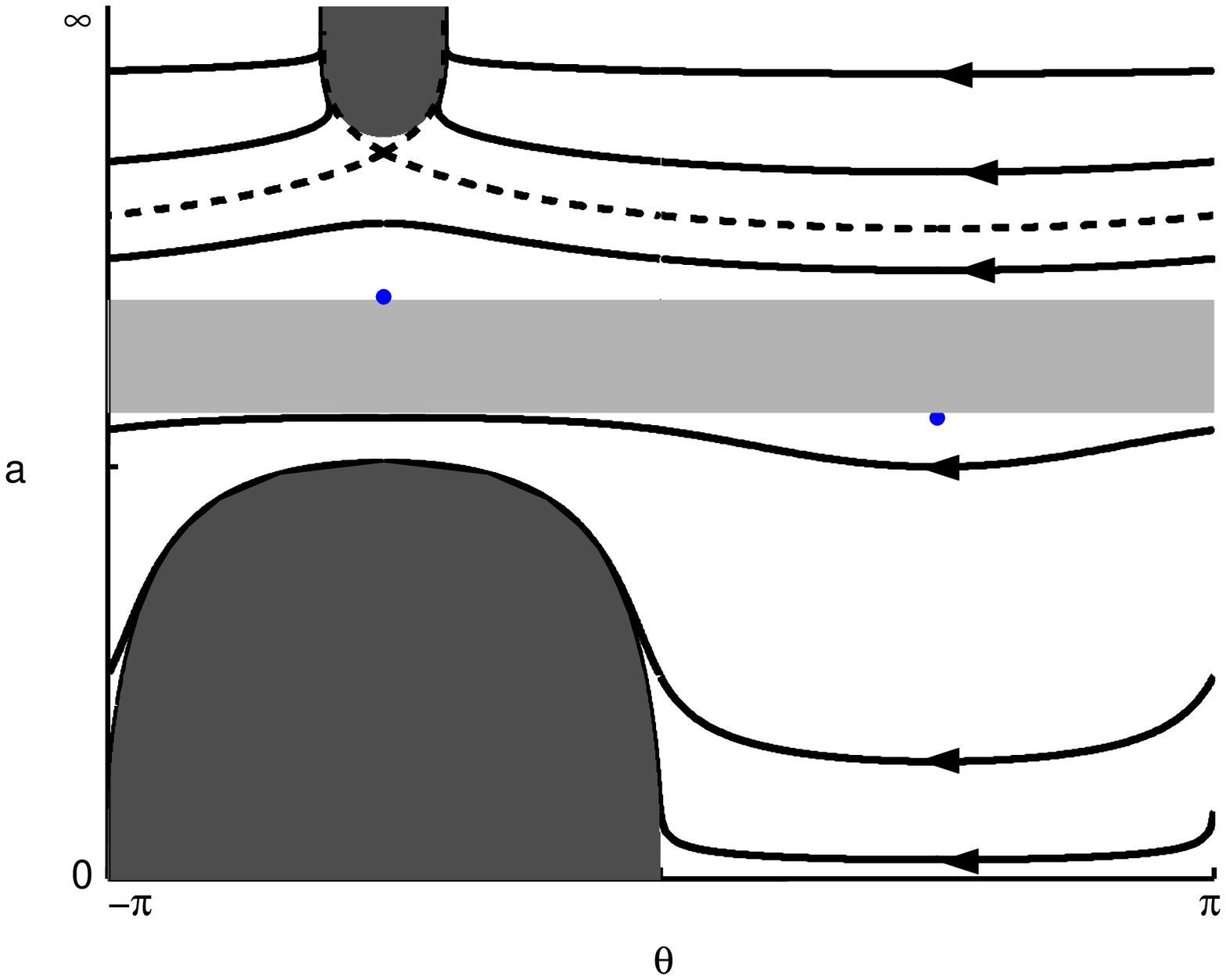}
\caption[] {\label{fig4}
The top panel represents the phase space in the variables
$\{H,\dot{\phi},a\}$, when
$m=1.0$, $\sigma=0.05$ and $0<V<V_{\rm crit2}$.  The
surface defined by the Friedmann equation (\ref{FriedSahni}), with
$\rho=\dot{\phi}^2/2+V$, is also plotted. The phase space
trajectories lie on this surface.  The axes have been compactified using
the rescalings $x={\rm arctan} (H)$, 
$y= {\rm arctan}(\ln \dot{\phi})$ and $z=
{\rm arctan}(\ln a)$. 
The lower panel illustrates a corresponding two-dimensional 
phase space. It follows from Eq. (\ref{FriedHy})
that there exists an axis of symmetry parallel to the $a$-axis
through the point $H=0$, $\dot{\phi}^2=2\sigma-2V$.
Cylindrical polar coordinates can then be defined by using this axis.
Specifically, we define $a=a$, ${\rm X}={\rm R} {\rm cos} \theta$ 
and ${\rm Y}= {\rm R} {\rm sin} \theta$,
where ${\rm X}=\frac{\sqrt{8 \pi \lpl^2}}{\sqrt{3}}\left
(\frac{\dot{\phi}^2}{2 \sqrt{2 \sigma}} - \left (
\frac{\sqrt{2 \sigma}}{2} -\frac{V}{\sqrt{2 \sigma}} \right ) \right )$,
${\rm Y}=H$, and ${\rm R}= \left ( \left (
\frac{\sqrt{m}}{a^2}-\frac{1}{2\sqrt{m}}\right)^2 + \frac{4 \pi \lpl^2 
\sigma}{3} - \frac{1}{4 m} \right)^{\frac{1}{2}}$. The two-dimensional plot
is then presented in the $\{\theta, a\}$ plane, where 
we have once more compactified the $a$-axis 
using $y={\rm arctan}(\ln a)$.  This diagram is
essentially a projection of the top panel about the axis defined above, 
with the points at $\theta = \pi$
identified with those at $\theta=-\pi$.  The shaded areas mark the
unphysical regions in this coordinate system.
The centre equilibrium points
are identified by a solid circle and the
separatrix is represented by a dotted line. The saddle point occurs
at the point where the separatrix intersects with itself.
}
\end{figure}

\begin{figure}[!t]
\includegraphics[width=8.1cm, height=6.6cm]{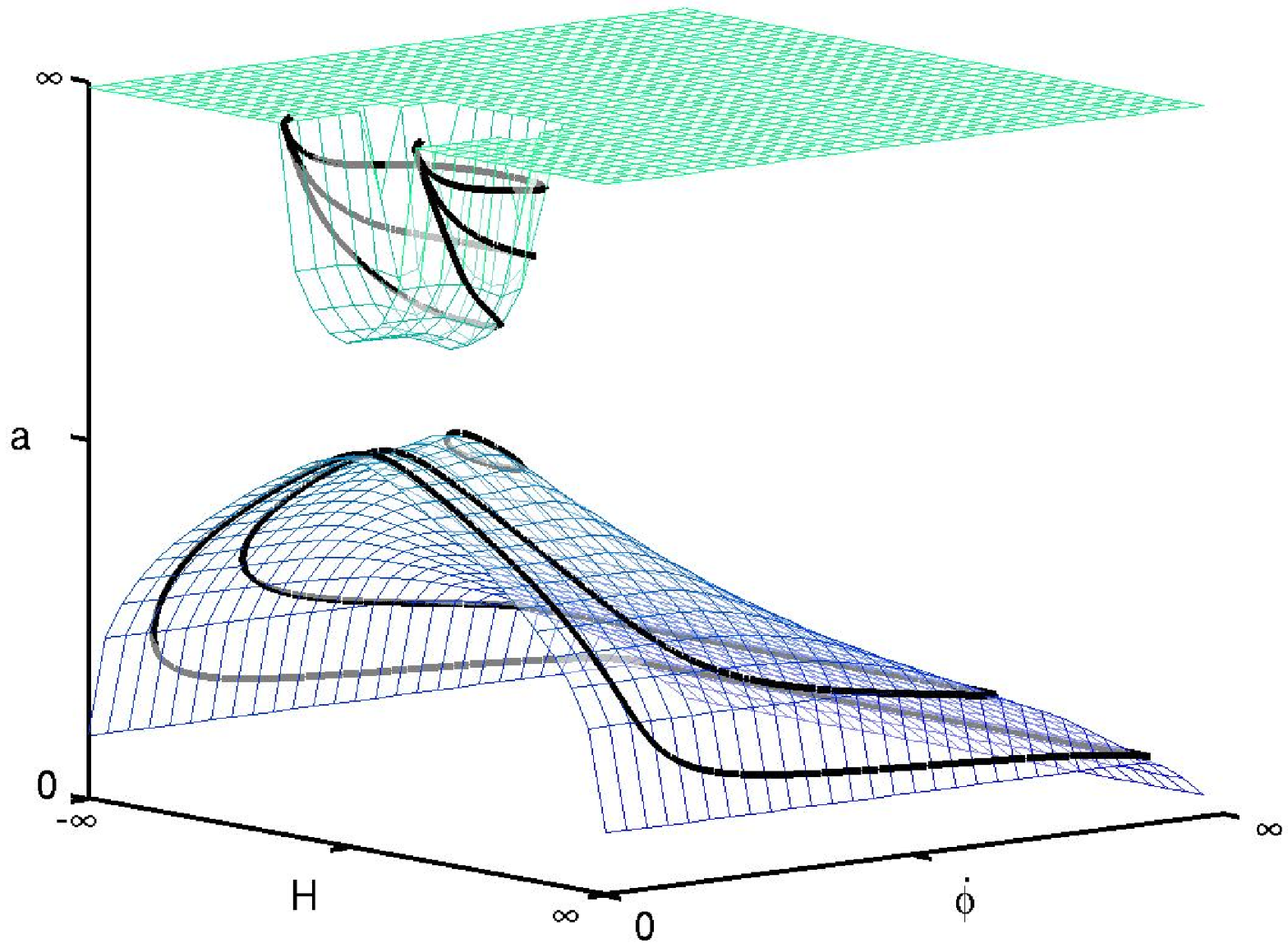}
\includegraphics[width=7.2cm]{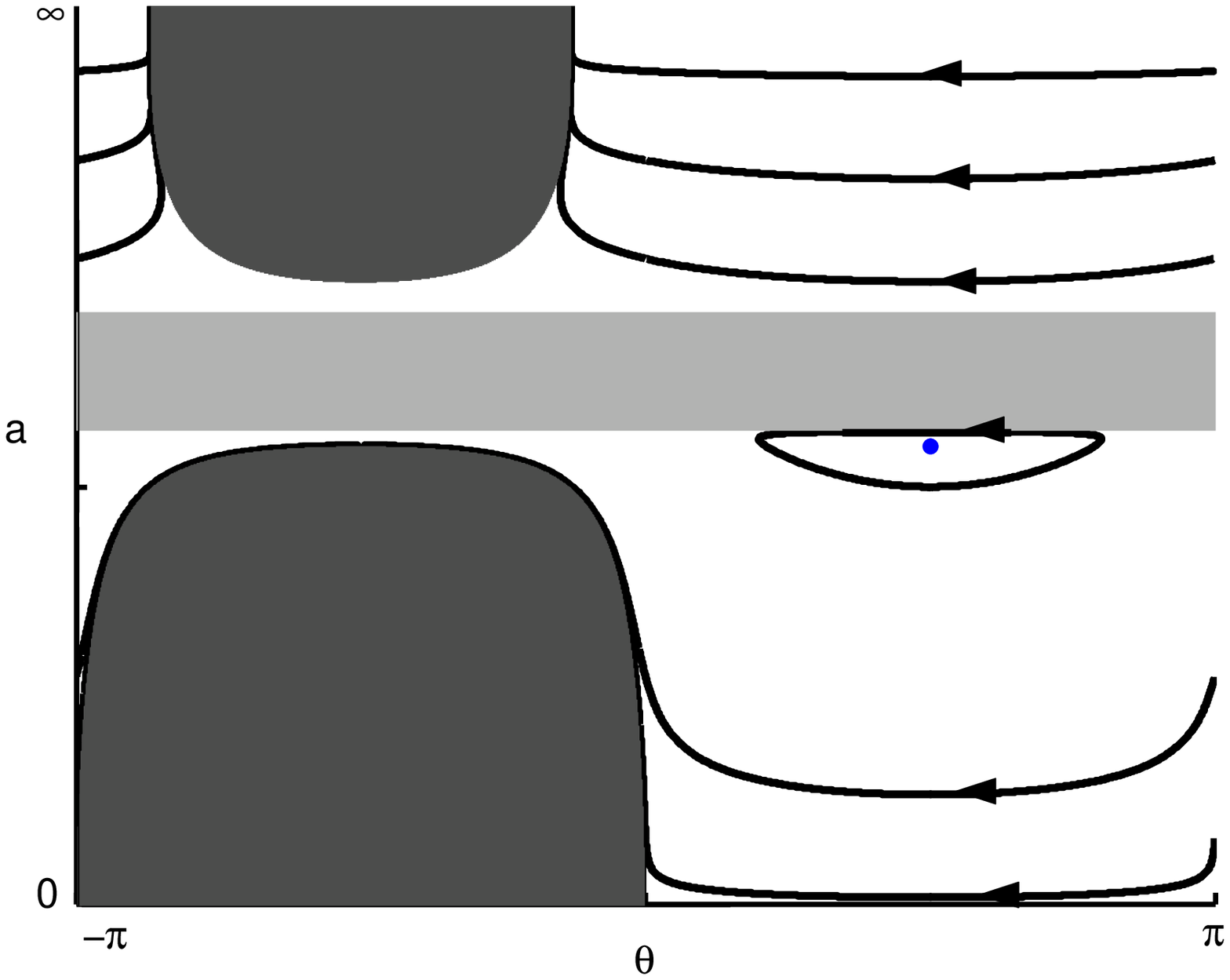}
\caption[] {\label{fig5}
As in Fig. \ref{fig4}, but now for a potential $V>V_{\rm crit2}$.  
There are no saddle points and only one
centre equilibrium point in the phase plane. 
}
\end{figure}

The qualitative evolution of the universe in the regimes 
$V<V_{\rm crit2}$ and $V>V_{\rm crit2}$ is  
shown in Figs. \ref{fig4} and \ref{fig5}, respectively, 
where the trajectories have been calculated by numerically 
integrating the field equations for the specific choices 
$m =1$ and $\sigma =0.05$.  
The topology of the Friedmann surface in the upper panels of 
these figures, together with Eq. (\ref{FriedHy}), implies that 
the entire phase space can be represented as a single two-dimensional 
plot by employing cylindrical 
polar coordinates. These are shown in the lower panels of the figures.

The dynamics in the upper sectors of the Friedmann surface 
in Figs. \ref{fig4} and \ref{fig5}
is qualitatively similar 
to the case where no dark radiation is present ($m=0$). The lower 
sector of the Friedmann surface contains the new centre equilibrium point 
that arises when $m \ne 0$. 
This point is real for all values of the potential. Hence, as shown in 
Figs. \ref{fig4} and \ref{fig5}, the universe remains trapped 
in an indefinite cycle of expansion and contraction if 
it is initially located in the lower half of the Friedmann surface.
This behaviour can be understood since a sufficiently large (positive) 
cosmological constant introduces a large negative 
{\em effective} cosmological constant in the 
Friedmann equation (\ref{FriedSahni}) as a consequence 
of the quadratic term in the energy density. 
This will prevent the universe from 
entering a phase of accelerated inflationary expansion. 
Indeed, although is is not relevant for the graceful entrance mechanism, 
it can be shown that if $V$ exceeds yet another critical value, 
the upper region of the Friedmann surface is no longer physical 
since the field's kinetic energy becomes negative. 
In effect, therefore, the size of the universe is bounded from above.  

By comparing the qualitative dynamics of the universe when 
no dark radiation is present (Figs. \ref{fig1} and \ref{fig2}) to that 
illustrated in the upper sectors of the Friedmann surfaces 
in Figs. \ref{fig4} and \ref{fig5}, we may infer that 
a graceful entrance to inflation, as outlined in Section II, 
may in principle occur for suitable choices of initial conditions. 


\subsection{$m > m_{\rm crit}$}

\begin{figure}[!t]
\includegraphics[width=8.2cm]{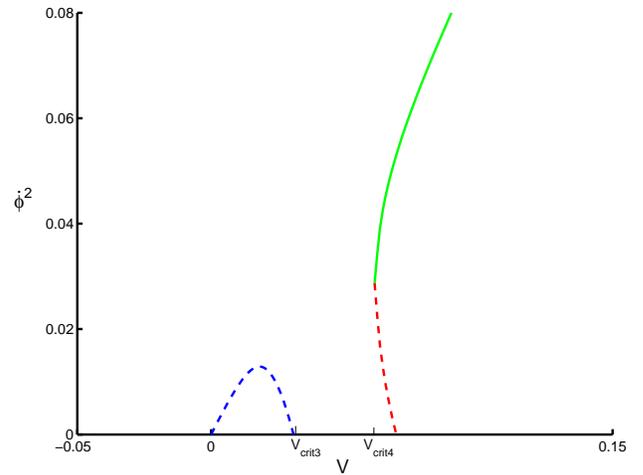}
\caption[] {\label{fig6}
As for Fig. \ref{fig3}, but now for the case where
$m>m_{\rm crit}$. Numerical values chosen for the parameters 
are $m=1.4$ and $\sigma=0.05$ in Planck units.
}
\end{figure}

We may adopt a similar approach for $m>m_{\rm crit}$. 
Fig. \ref{fig6} illustrates the position and nature of the physically 
relevant equilibrium points as a function of the field's kinetic 
and potential energies for the specific choices 
$m=1.4$ and $\sigma =0.05$. The qualitative nature of this plot 
remains unaltered for all $m> m_{\rm crit}$. It follows from 
Fig. \ref{fig6} that there is only a single saddle equilibrium point
when the magnitude of the potential is below a critical value 
$V_{\rm crit3}$. There is then a finite 
range of values of the potential, $V_{\rm crit3} \le  V \le 
V_{\rm crit4}$, for which there are {\em no} physical equilibrium points. 
For $V>V_{\rm crit4}$, on the other hand,  
there exists both a centre and a saddle equilibrium point, 
and at still higher values of the potential, the saddle point 
disappears once more. However, the equilibrium points which occur 
above $V_{\rm crit4}$ are not relevant to the graceful entrance scenario.

\begin{figure}[!t]
\includegraphics[width=8.05cm, height=6.6cm]{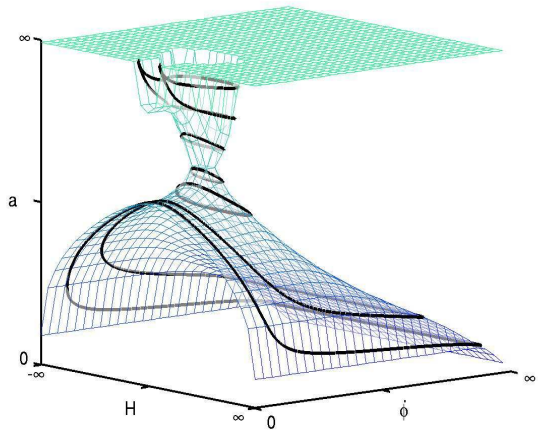}
\includegraphics[width=7.2cm]{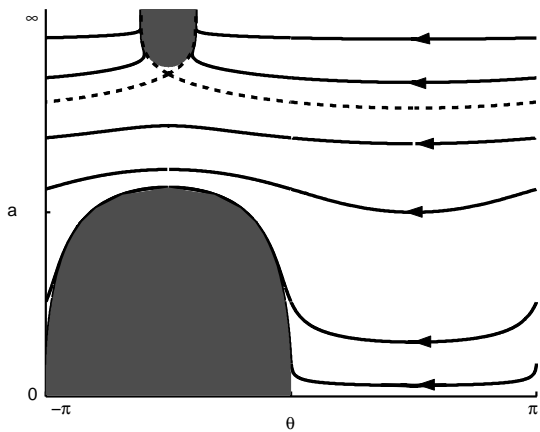}
\caption[] {\label{fig7}
The top and bottom panels are as described for Fig. \ref{fig4},
but now for $m>m_{\rm crit}$ and $0<V<V_{\rm crit3}$.  
Numerical values for the parameters are
$m=1.4$ and $\sigma=0.05$ in Planck units.  Trajectories 
representing cyclic behaviour arise in this case 
but there are no centre equilibrium points and only a single 
saddle point is present.
}
\end{figure}

\begin{figure}[!t]
\includegraphics[width=8.05cm, height=6.6cm]{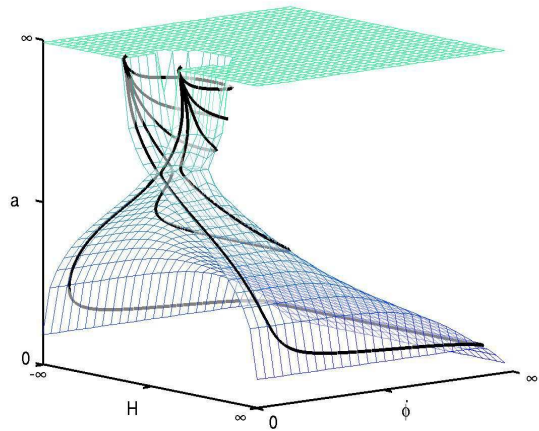}
\includegraphics[width=7.2cm]{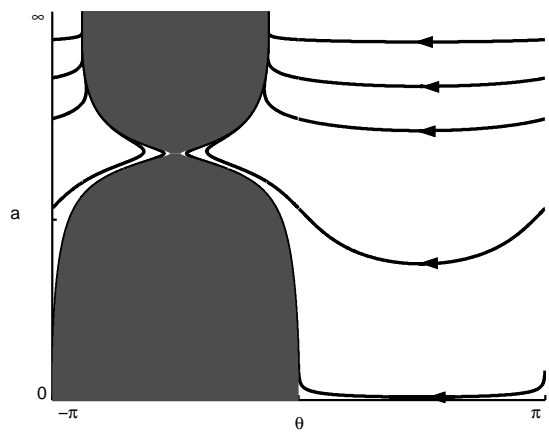}
\caption[] {\label{fig8}
The top and bottom panels are as described for Fig. \ref{fig4},
but now for $m>m_{\rm crit}$ and $V_{\rm crit3}<V<V_{\rm crit4}$.  
Numerical values of the parameters are 
specified as $m=1.4$ and $\sigma=0.05$ in Planck units.  
There are no equilibrium points for this case. 
}
\end{figure}

Figs. \ref{fig7} and \ref{fig8} illustrate the Friedmann surface and 
and the corresponding phase trajectories for $V< V_{\rm crit3}$
and $V_{\rm crit3} < V < V_{\rm crit4}$, respectively. 
In Fig. \ref{fig7}, the saddle point represents a divide between 
cyclic and inflationary behaviour, although it should 
be noted that the cyclic dynamics does not occur around 
a centre equilibrium point in this case.  For $V >V_{\rm crit3}$,  
the saddle point disappears and all trajectories eventually 
evolve into an inflationary region, as shown in  Fig. \ref{fig8}.  
As was the case in the previous subsection, it can be shown 
that for a sufficiently large value of the potential, 
the Friedmann surface becomes modified in such a way that the 
surface is bounded to always lie below a critical value of the scale 
factor. This implies that the size of the universe is bounded 
from above and such behaviour follows once more due 
to the presence of an effective negative cosmological 
constant in the Friedmann equation. 


The question of whether a graceful entrance to inflation 
is possible if $m>m_{\rm crit}$ is more difficult to answer.  
The key features of a cyclic region and a saddle point in the 
phase space, which disappear at a critical value of the potential, 
are indeed present. However, there is a further complication. For 
$m<m_{\rm crit}$, the cyclic dynamics is always dominated 
by the field's kinetic energy, but this is no 
longer the case when $m>m_{\rm crit}$. 

Further insight may be gained from Figs. \ref{fig7} and \ref{fig8}.  
The shaded gray areas (red in the online versions) represent the regions 
of phase space where the field's kinetic energy is negative, 
with the boundaries corresponding to the limit $\dot{\phi}^2=0$. 
For a realistic (i.e. sufficiently flat but field-dependent) potential, 
this implies that on a trajectory which passes sufficiently 
close to these boundaries in the `instantaneous phase space', 
the field's kinetic energy will be so small that the gradient of the 
potential will become significant. This will result in 
the kinetic energy of the field falling to zero and the field turning around
on the potential before it has climbed sufficiently far up the potential 
to drive a successful phase of slow-roll inflation. 
In the previous cases, such behaviour was only possible 
in the region of parameter space relevant to inflation, or 
in a region disconnected from it, such as the lower 
section of the Friedmann surface. In the present case, however, 
the two regions of the previously disjointed Friedmann surface are 
connected. In effect, therefore, the 
turn around in the field may occur too soon for inflation 
to occur if $m>m_{\rm crit}$. 

More specifically, let us assume as before that we have a 
realistic inflationary potential, with a magnitude initially 
in the range $0< V<V_{\rm crit3}$, 
and that the universe begins in the oscillatory region of the phase space. 
As the universe oscillates, the field climbs its potential. 
In terms of Fig. \ref{fig7}, the instantaneous phase space is 
therefore altered, with the saddle point moving to successively 
smaller values of the scale factor, and the shaded (physically forbidden) 
regions growing in size. Ultimately, a point will be attained 
when the universe's trajectory either moves outside the region of the 
oscillations or it passes too 
close to the boundary of the lower, shaded region. If the former 
behaviour arises, the graceful entrance dynamics applies as in 
the previous scenarios discussed above. On the other hand, in the latter 
case, the field may turn around on the potential 
{\it without} inflation occurring, 
since the scale factor will still be small at this stage 
and consequently the curvature 
term in the Friedmann equation will be significant. 
 Such behaviour is qualitatively 
similar to the effect discussed in the LQC scenario 
when perfect fluid matter sources are introduced into the system 
\cite{nunes}.  
The issue of whether a graceful entrance to inflation may occur is therefore 
sensitive to the initial conditions.  

\section{Discussion}

In this paper, we have examined the criteria that a 
cosmological model must satisfy in order for it 
to undergo a `graceful entrance' into a phase of inflation, whereby
a scalar field is able to move 
up its interaction potential whilst the universe undergoes 
a large number of oscillatory cycles.  
Our approach has been to consider an idealized model, 
where the matter is comprised of a scalar field evolving 
along a constant potential. This system 
provides a good approximation to more realistic 
scenarios where the potential is field-dependent 
if the change in the potential is sufficiently small over a 
large number of cycles. It therefore provides us with a methodology 
for identifying whether a particular 
cosmological scenario will meet the necessary 
requirements for graceful entrance. 

The mechanism we have outlined is very generic.
The important ingredients are that the phase space 
should exhibit both a centre and saddle equilibrium point
when the magnitude of the potential $V$ falls below a critical value 
$V_{\rm crit}$. These points gradually move towards each other 
as $V$ increases and eventually merge when $V=V_{\rm crit}$. 
Above this scale, the points should disappear from 
the phase space. From the physical viewpoint, this behaviour arises  
because the potential now dominates the field's 
kinetic energy, $V > \dot{\phi}^2$, and consequently drives a 
phase of inflationary expansion. It follows that the critical 
value $V_{\rm crit}$ sets the energy scale for the onset of inflation.  

We have developed a framework for investigating 
a general class of braneworld models in this context that are 
characterized by a Friedmann equation with an arbitrary 
dependence on the energy density. Such a class of models 
can be represented as a standard, relativistic 
Friedmann universe by reinterpreting the effective 
equation of state of the matter on the brane. 
As a concrete example, we focused on the 
Shtanov-Sahni braneworld \cite{shsh}, where the extra bulk dimension is 
timelike. We found that the question of whether the 
braneworld oscillates or expands indefinitely 
is dependent on the field's kinetic and potential energies, as well 
as the dark radiation and the brane tension, $\sigma$. If the 
dark radiation is negligible, the phase space has a similar structure 
to that outlined above and a graceful entrance to inflation 
can therefore be realised, where the energy scale 
at the onset of inflation is given by $V_{\rm crit} \approx \sigma$.
The dark radiation can have a significant effect on the dynamics, however, 
and the question of whether a graceful entrance is possible 
when this term is present is sensitive to the initial conditions. 

In conclusion, therefore, we have presented a non-singular, 
oscillating braneworld that can in principle exhibit
a graceful entrance to inflation. 
We anticipate that such behaviour will apply for a wide range of 
collapsing braneworld scenarios that are able to undergo non-singular 
bounces (see, e.g., \cite{burgess}) 
and, moreover, will not be strongly dependent on the precise form 
of the inflaton potential. Similar dynamics 
was recently shown to arise in models inspired by 
loop quantum cosmology \cite{lmnt,mntl,mtle} and it is interesting 
that such behaviour is also possible 
in braneworld models motivated by string/M-theory.
 
Finally, our analysis also provides the basis for realising 
the recently proposed emergent universe scenario in a braneworld 
context \cite{em,mtle}. In this scenario, 
the initial state of the universe is postulated to be 
either an Einstein static universe or one that is oscillating about such a 
solution in the infinite past. The potential 
has an asymptotically flat section, where 
$V<V_{\rm crit}$ as $\phi  \to -\infty$, but subsequently increases above
$V_{\rm crit}$ once the value of the field has increased beyond a 
certain value. If the field is initially located in the plateau region
of the potential and moving in the appropriate direction, it will eventually 
reach the section of the potential which rises above $V_{\rm crit}$, 
thus initiating inflation. Within the framework of classical 
Einstein gravity, the emergent universe scenario suffers from severe
fine-tuning, since the static universe
corresponds to a saddle point in the phase space. 
However, a mechanism which allows for a graceful entrance to inflation
provides a natural realization of this scenario 
with less fine tuning and a greater degree of freedom in the 
choice of inflationary potential. 

\section*{Acknowledgments}  
DJM is supported by PPARC. We thank R. Tavakol for numerous 
discussions.

\end{document}